\documentclass[prb,aps,twocolumn,showpacs,superscriptaddress,nobibnotes,epsf]{revtex4}
\usepackage{graphicx}% Include figure files
\usepackage{dcolumn}% Align table columns on decimal point
\usepackage{bm}% bold math
\usepackage{SIunits}
\usepackage{tabularx}

\begin{document}

\title{Calorimetric study in single crystal of CsFe$_2$As$_2$}

\author{A. F. Wang}
\affiliation{Hefei National Laboratory for Physical Science at Microscale and Department of
Physics, University of Science and Technology of China, Hefei, Anhui 230026, People's Republic
of China}

\author{B. Y. Pan}
\affiliation{Department of Physics, State Key Laboratory of Surface
Physics, and Laboratory of Advanced Materials, Fudan University,
Shanghai 200433, China}

\author{X. G. Luo}
\affiliation{Hefei National Laboratory for Physical Science at Microscale and Department of
Physics, University of Science and Technology of China, Hefei, Anhui 230026, People's Republic
of China}

\author{F. Chen}
\affiliation{Hefei National Laboratory for Physical Science at
Microscale and Department of Physics, University of Science and
Technology of China, Hefei, Anhui 230026, People's Republic of
China}

\author{Y. J. Yan}
\affiliation{Hefei National Laboratory for Physical Science at
Microscale and Department of Physics, University of Science and
Technology of China, Hefei, Anhui 230026, People's Republic of
China}

\author{J. J. Ying}
\affiliation{Hefei National Laboratory for Physical Science at
Microscale and Department of Physics, University of Science and
Technology of China, Hefei, Anhui 230026, People's Republic of
China}

\author{G. J. Ye}
\affiliation{Hefei National Laboratory for Physical Science at
Microscale and Department of Physics, University of Science and
Technology of China, Hefei, Anhui 230026, People's Republic of
China}

\author{P. Cheng}
\affiliation{Hefei National Laboratory for Physical Science at
Microscale and Department of Physics, University of Science and
Technology of China, Hefei, Anhui 230026, People's Republic of
China}

\author{X. C. Hong}
\affiliation{Department of Physics, State Key Laboratory of Surface
Physics, and Laboratory of Advanced Materials, Fudan University,
Shanghai 200433, China}

\author{S. Y. Li}
\affiliation{Department of Physics, State Key Laboratory of Surface
Physics, and Laboratory of Advanced Materials, Fudan University,
Shanghai 200433, China}

\author{X. H. Chen}
\altaffiliation{Corresponding author} \email{chenxh@ustc.edu.cn}
\affiliation{Hefei National Laboratory for Physical Science at Microscale and Department of
Physics, University of Science and Technology of China, Hefei, Anhui 230026, People's Republic
of China}

\begin{abstract}
We measured resistivity and specific heat of high-quality CsFe$_2$As$_2$ single crystals, which were grown by using a
self-flux method. The CsFe$_2$As$_2$ crystal shows sharp
superconducting transition at 1.8 K with the transition width of 0.1
K. The sharp superconducting transition and pronounced jump in
specific heat indicate high quality of the crystals. Analysis on the
superconducting-state specific heat supports unconventional pairing
symmetry in CsFe$_2$As$_2$.

\end{abstract}


\vskip 300 pt

\maketitle

\section{INTRODUCTION}

The discovery of iron-based superconductors has opened a new window
for unveiling the physics of high-temperature superconductivity
besides cuprates.\cite{Hosono,Chen,BaK} Among the various families
of iron-based superconductors discovered till now,
$A$Fe$_2$As$_2$($A$=alkali earth, alkali, and Eu, the so called 122
system), which has the ThCr$_2$Si$_2$ structure, was the most
investigated due to the easy growth of sizable high-quality single
crystals. \cite{Wang XF PRL} In this 122 system, KFe$_2$As$_2$, as
the end member of the Ba$_{1-x}$K$_x$Fe$_2$As$_2$ series, shows some
unique properties. Firstly, superconductivity with $T_c$ $\sim$ 4 K
can be realized in KFe$_2$As$_2$ without purposely doping. Secondly,
very clean single crystal with residual resistivity ratio(RRR)
exceeding 1000 can be quite easily achieved, which is a good start
point to study intrinsic physical properties. It was proposed that
inter-band interaction that links the hole and electron Fermi
surfaces (FS) produces an $\emph{s}_{\pm}$ pairing symmetry in most
of the iron-based superconductors. However, angle-resolved
photoemission spectroscopy (ARPES) and the de Hass-van Alphen (dHvA)
experiments revealed that the electron pockets disappeared  and the
large hole sheets centered around $\Gamma$ point dominate FS in
KFe$_2$As$_2$. \cite{Ding H, Haas Alphen} Therefore, the pairing
interaction could be distinct from other iron-based superconductors.
The nodes on superconducting gaps have been detected by thermal
conductivity, \cite{Li SY, Louis Taillefer} penetration depth,
\cite{penetration depth} and NMR. \cite{NMR, Yu WQ} The measurements
of thermal conductivity, \cite{Wang AF, Louis Taillefer} specific
heat \cite{KNa} and penetration depth \cite{penetration depth}
support a \emph{d}-wave superconducting state in KFe$_2$As$_2$. In
contrast, recent ultrahigh-resolution laser ARPES suggests a nodal
\emph{s}-wave superconductor with highly unusual FS-selective
multi-gap structure.\cite{Science} Whether those nodes are imposed
by symmetry or accidental still remains an open question. As a
consequence, further investigation on analogous compounds would be
significant to clarify the underlying physics in $A$Fe$_2$As$_2$
system. In the present article, we will report the crystal growth
and characterization of the analogous compound CsFe$_2$As$_2$.

Superconductivity at 2.6 K was observed in the polycrystalline
CsFe$_2$As$_2$ sample. \cite{Chu CW} But few physical properties
have been reported so far because high quality single crystals are
not available until now.  The difficulty of growing sizable
CsFe$_2$As$_2$ single crystal mainly lies in the extremely high
chemical activity and low melting point of Cs. In the present
article, we have successfully overcome this problem by using the
stainless steel sample container assembly, \cite{Hiroshi} which can
be sealed in the glove box (O$_2$ content is less than 1 ppm)
mechanically. As a result, sizable high-quality single crystals of
CsFe$_2$As$_2$ were grown. The CsFe$_2$As$_2$ single crystals were
characterized by X-ray diffraction (XRD), resistivity, magnetic
susceptibility, and specific heat. The sharp superconducting
transition temperature and obvious specific jump indicate good
quality of the single crystals.

\section{EXPERIMENTAL DETAILS}

  High quality CsFe$_2$As$_2$ single crystals are grown by the self flux
technique. The Cs chunks, Fe and As powder were weighted according
to the ratio Cs:Fe:As=6:1:6. Typically, 1.5 grams of the mixture of
Fe and As powders were loaded into a 10 mm diameter alumina
crucible, and freshly cut Cs pieces were placed on top of the
mixture. Then the alumina crucible with a lid was sealed in a
stainless steel container assembly. The whole preparation process
was carried out in the glove box in which high pure argon atmosphere
is filled (O$_2$ content is less than 1 ppm). Considering the low
melting point of Cs ($T_{\rm m}$=28 $^{\circ}$C), the room
temperature must be kept blow 20 $^{\circ}$C. The sealed stainless
steel assembly was then sealed inside an evacuated quartz tube. The
quartz tube was placed in a box furnace and slowly heated up to 200
$^{\circ}$C, It was kept at 200 $^{\circ}$C for 400 minutes, which
allows full reaction of Cs and the mixture. Then the sample was
heated up to 950 $^{\circ}$C in 10 hours. The temperature was kept
still for 10 hours and then slowly cooled to 550 $^{\circ}$C at a
rate of 3 $^{\circ}$C/h. After cooling down to room temperature by
switching off the furnace, shiny plate-like crystals can be easily
picked up from the alumina crucible. The single crystals are stable
in air or alcohol for several days.

XRD was performed on a SmartLab-9 diffracmeter(Rikagu) from 10$^{\rm
o}$ to 80$^{\rm o}$ with a scanning rate of 4$^{\rm o}$ per minute.
The actual chemical composition of the single crystal is determined
by energy dispersive x-ray spectroscopy(EDX) mounted on the field
emission scanning electronic microscope (FESEM), Sirion 200.
Magnetic susceptibility was measured using Vibrating Sample
Magnetometer (VSM) (Quantum Design). The direct current (dc)
resistivity was measured by conventional four probe method using the
PPMS-9T (Quantum Design). Resistivity and specific heat down to 50
mK were measured in a dilution refrigerator on PPMS.

\begin{figure}[ht]
\centering
\includegraphics[width=0.49\textwidth]{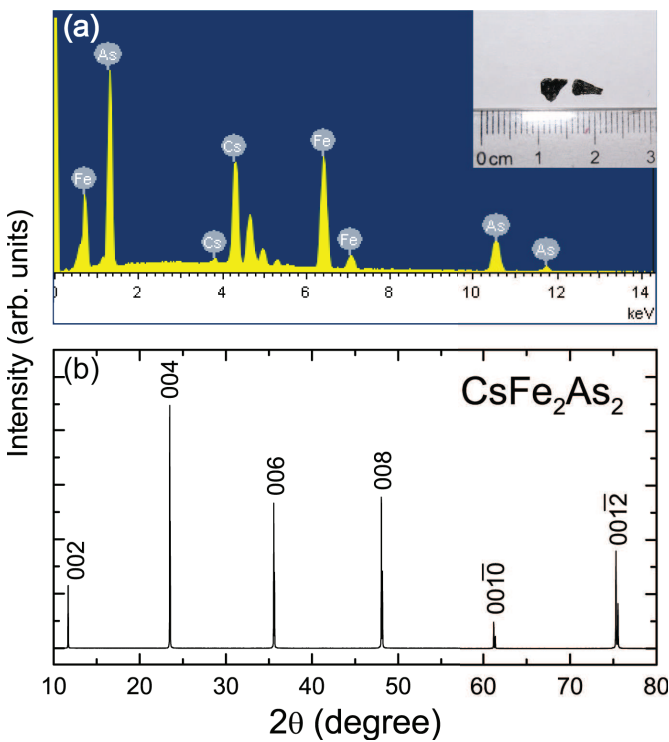}
\caption{(color online). (a) The typical EDX spectrum for
CsFe$_2$As$_2$ single crystal, the inset shows photograph of the
CsFe$_2$As$_2$ single crystal together with a millimeter scale.  (b)
X-ray diffraction pattern of the CsFe$_2$As$_2$ single crystal.}
\label{fig1}
\end{figure}

\section{RESULTS}

The typical size of as grown single crystals is about 5 mm $\times$3
mm $\times$ 0.03 mm, as shown in the inset of Fig. 1(a). Elemental
analysis was performed using EDX. A typical EDX spectrum is shown in
Fig.1(a), and the obtained atomic ratio of Cs:Fe:As is roughly
20.76: 40.52: 38.71. The atomic ratio is consistent with the
composition CsFe$_2$As$_2$ within instrumental error. Fig.1(b) shows
the single crystal XRD pattern for CsFe$_2$As$_2$. Only (00\emph{l})
reflections can be recognized, indicating that the crystal is well
orientated along the \emph{c} axis. The \emph{c}-axis lattice
parameter was estimated to be \emph{c}=15.13 {\AA}, consistent with
previous report on the polycrystalline samples. \cite{Chu CW}

\begin{figure}[ht]
\centering
\includegraphics[width=0.49\textwidth]{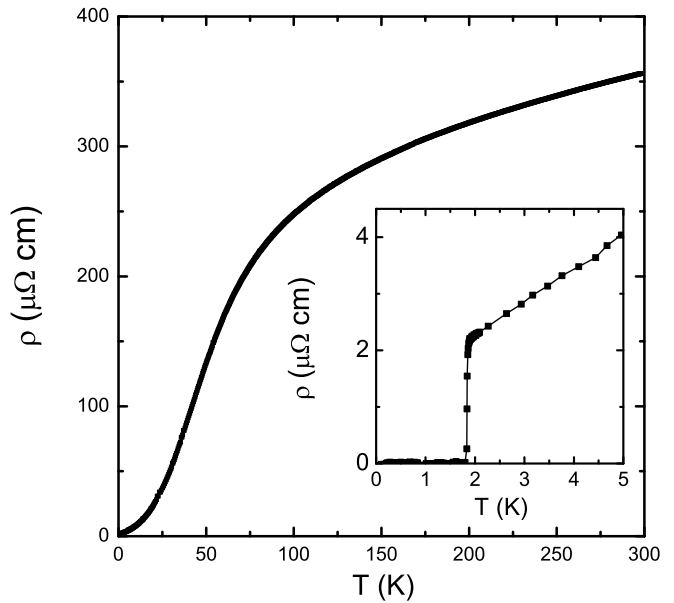}
\caption{(color online). Resistivity plotted as a function of temperature
for CsFe$_2$As$_2$ single crystal. The inset is the zoom plot of
resistivity around the superconducting transition.} \label{fig2}
\end{figure}

Fig. 2 shows the in-plane resistivity as the function of the
temperature for CsFe$_2$As$_2$ single crystal. The resistivity
exhibits metallic behavior in the entire temperature range between
60 mK to 300 K. The behavior of the resistivity resembles the
counterpart compound KFe$_2$As$_2$. The resistivity begins to drop
rapidly at 1.88 K and reaches zero at 1.8 K. The superconducting
transition temperature $T_{\rm c}$ in the following text is defined
as the temperature when the resistivity reaches zero. The sharp
superconducting transition with transition width less than 0.1 K
indicates high quality of the crystal. The $T_{\rm c}$ of
CsFe$_2$As$_2$ is slightly lower than that reported on the
polycrystalline sample. \cite{Chu CW} The residual resistivity ratio
(RRR) $\rho$(300K)/$\rho$(5K) is estimated to be 88, comparable with
those in the crystals of KFe$_2$As$_2$ grown with FeAs flux,\cite{Li
SY, Wang NL} but much smaller than those in crystals of
KFe$_2$As$_2$ grown with KAs. \cite{Wang AF, Hiroshi}

\begin{figure}[ht]
\centering
\includegraphics[width=0.49\textwidth]{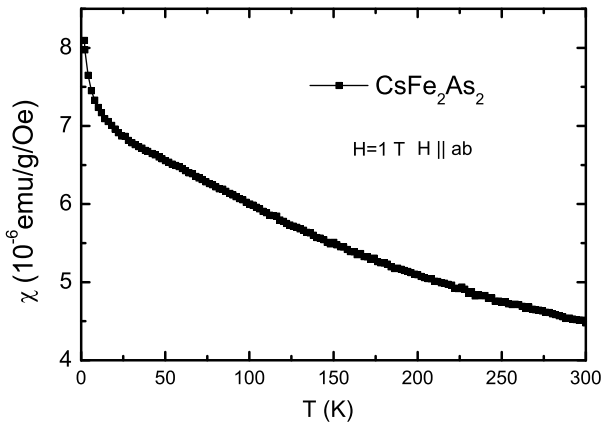}
\caption{(color online). Temperature dependence of magnetic
susceptibility of the CsFe$_2$As$_2$ single crystal collected at $H$=1 T with $H$
$\parallel$ $ab$.} \label{fig3}
\end{figure}

Fig. 3 shows the the temperature dependence of the in-plane magnetic
susceptibility for CsFe$_2$As$_2$ single crystal under H = 1 T in
the normal state. The CsFe$_2$As$_2$ single crystal shows
paramagnetic behavior from 300 K to 2 K, and no magnetic anomaly was
observed. The magnetic susceptibility behavior of CsFe$_2$As$_2$ is
similar to KFe$_2$As$_2$. \cite{magnetic susceptibility}

\begin{figure}[ht]
\centering
\includegraphics[width=0.49\textwidth]{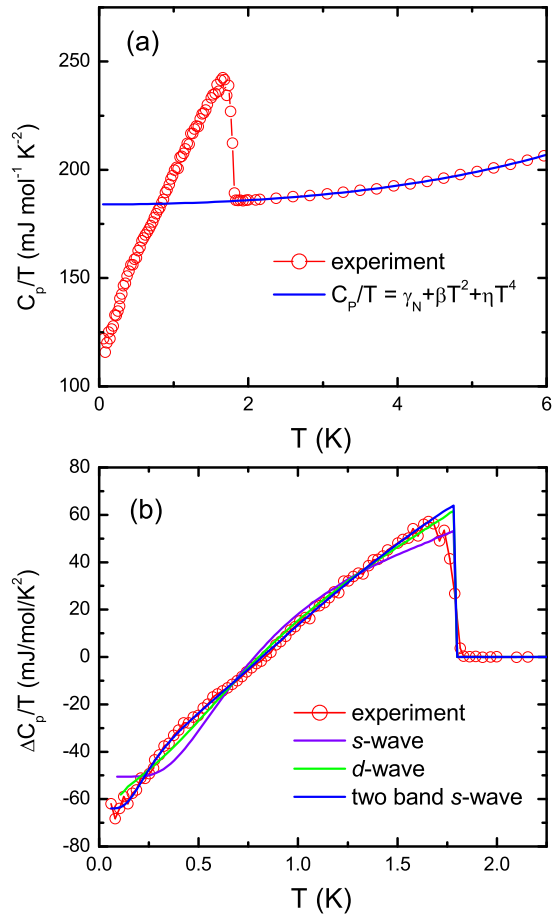}
\caption{(color online). (a) Temperature dependence of specific heat
divided by temperature $C_{\rm p}(T)/T$ of CsFe$_2$As$_2$.
The blue solid line is the fitting for the data between
1.9 and 10 K (data below 6 K are shown). The fitting function is
described in the text. (b) Temperature dependence of the difference
in the electronic specific heat between the superconducting state
and the normal state. The violet, green and blue solid line
represent fits for the experimental data by using single band \emph{s}-wave ($\alpha$ = 1.42), single band \emph{d}-wave ($\alpha$ = 1.93) and
two band \emph{s}-wave $\alpha$ model ($\alpha_1$ =1.21, $\alpha_2$ =3.33), respectively.} \label{fig4}
\end{figure}

In Fig. 4(a), we show the temperature dependence of the specific
heat of CsFe$_2$As$_2$ down to 50 mK under zero magnetic field. A
pronounced jump due to the superconducting transition is observed
below 1.8 K, consistent with the resistivity measurement. This
indicates the high quality of the present crystals. The normal-state
specific heat can be well fitted by
$C_{\rm{normal}}(T)$=$\gamma_{\rm N}T$+$C_{\rm{lattice}}(T)$, where
$\gamma_{\rm N}T$ and $C_{\rm{lattice}}(T)$=$\beta T^3$+$\eta T^5$
are electron and phonon contributions, respectively.\cite{Wang AF Na} The solid line
in Fig. 4(a) is the best fit to the $C_{\rm p}/T$ above $T_{\rm c}$
(1.9 K to 10 K). We obtained $\gamma_{\rm N}$=184.01 mJ
$\rm{mol}^{-1}$ $\rm{K}^{-2}$, $\beta$=0.466 mJ $\rm{mol}^{-1}$
$\rm{K}^{-4}$ and $\eta$=0.00474 mJ $\rm{mol}^{-1}$ $\rm{K}^{-6}$.
From this value of $\beta$ and by using the formula ${\Theta}_{\rm
D}$=[12$\pi^4$$k_B$$N_A$$Z$/(5$\beta$$)]^{1/3}$, where $N_{\rm A}$
is the Avogadro constant and $Z$ is the total number of atoms in one
unit cell, the Debye temperature (${\Theta}_{\rm D}$) is estimated
to be 275 K. This value is comparable to that of KFe$_2$As$_2$.
\cite{Canfield} It should be pointed out that $\gamma$$_{\rm
N}$=184.01 mJ $\rm{mol}^{-1}$ $\rm{K}^{-2}$ is very large. The
specific heat jump is only about 31\% of $\gamma$$_{\rm N}$, which
is similar to the case of KFe$_2$As$_2$. \cite{NMR, KNa, Canfield,
magnetic susceptibility} Considering the sharp superconducting
transition, this feature may be the evidence for the existence of
low energy quasiparticle excitation. \cite{NMR} $\Delta$$C_{\rm
P}$/$\gamma$$T_{\rm c}$ can be estimated to be about 0.35. This is
much smaller than the value (=1.43) expected for BCS
superconductors, which could indicate an unconventional pairing
symmetry. In addition, $\Delta$$C_{\rm P}$/$T_{\rm c}$ at $T_c$ =
1.8 K is estimated to be about 64 mJ mol$^{-1}$ K$^{-2}$, far above
the $\Delta$$C_{\rm P}$/$T_{\rm c}$  vs. $T_c$ behavior of other
iron-based superconductors, except for that observed in
KFe$_2$As$_2$. In KFe$_2$As$_2$, $\Delta$$C_{\rm P}$/$T_{\rm c}$
$\approx$ 41 mJ mol$^{-1}$ K$^{-2}$ at $T_c$ = 3.1 K, \cite{BNC}
which is also much above the $\Delta$$C_{\rm P}$/$T_{\rm c}$  vs.
$T_c$ behavior of other iron-based superconductors. This suggests
that the pairing symmetry in CsFe$_2$As$_2$ could be similar to that
in KFe$_2$As$_2$, but different from that of other iron-based
superconductors.

Fig. 4(b) shows the superconducting electronic specific heat
difference between the superconducting and normal states
$\Delta$$C_{\rm p}(T)$=$C_{\rm es}$$-$${\gamma}_{\rm N}T$=$C_{\rm
p}(T)$-$C_{\rm{normal}}(T)$, for which the entropy conservation is
confirmed to be satisfied. In the low-temperature limit, the
$\emph{s}$-wave model predicts
$\Delta$$C/T$$\cong$$a$$T^{-5/2}$exp$(-\Delta/{k_{\rm
B}T})$$-$$\gamma_{\rm n}$. While in clean $\emph{d}$-wave model,
$C_{\rm es}$$\sim$$T^2$, which gives $\Delta$$C_{\rm
p}/T$$=$$\alpha$$T$$-$${\gamma}_{\rm n}$, as have been observed in
organic superconductors $\kappa$-(BEDT-TTF)$_2$Cu[N(CN)$_2$]Br and
$\kappa$-(BEDT-TTF)$_2$Cu(NCS)$_2$. \cite{d wave} Fig. 4(b)
indicates that $\Delta$$C_{\rm p}/T$ can't be described by a
single-band BCS $\alpha$-model with $\alpha$=$\Delta$(0)/$k_{\rm
B}$$T_{\rm c}$=1.42. \cite{a model} As a consequence, our results
exclude the single-band $\emph{s}$-wave pairing symmetry. As shown
in Fig. 4(b), single-band $\emph{d}$-wave and two-band \emph{s}-wave
model give rise to better fitting quality. \cite{d wave2, Wang AF
Na} The low-temperature part of the specific heat roughly follows
linear temperature dependence rather than exponent, so the gap
symmetry is likely $\emph{d}$-wave. The thermal conductivity
measured on the same batch of CsFe$_2$As$_2$ single crystals
indicates nodal superconducting gap symmetry.\cite{LiSYC} It has
been reported that there is an anomaly around 0.7 K in $C_{\rm p}/T$
of KFe$_2$As$_2$, which was thought to be related to the impurity
contribution in the sample, \cite{Wang XF} and higher quality of
crystals can almost eliminate such anomaly. \cite{KNa} A very weak
anomaly-like feature is observed below 0.55 K in our data too, which
may affect the result of the fit. This makes it hard to make sure
the pairing symmetry from our fittings, since the main difference
between the fitting in $d$-wave and two-band $s$-wave models lies
below 0.55 K. Further measurements, such as ARPES and NMR, should be
performed to illustrate the pairing symmetry.

\section{SUMMARY AND CONCLUSIONS}

  We have successfully grown CsFe$_2$As$_2$ single crystals using the
assembly of stainless steel container. The sharp drop in resistivity
and pronounced jump in specific heat with $T_c$ around 1.8 K
indicate the high quality of crystals. The behavior of resistivity,
magnetic susceptibility, and specific data resemble the counterpart
compound KFe$_2$As$_2$. The $T$ dependence of the specific
heat in the superconducting state may be explained by a $d$-wave or a multi-band $s$-wave superconducting gap pictures. More work is required to reach a consensus for the pairing symmetry in this system.

\section*{Acknowledgements}
  This work is supported by the National Natural Science Foundation of China
(Grants No. 11190021, 11174266, 51021091), the "Strategic Priority
Research Program (B)" of the Chinese Academy of Sciences (Grant No.
XDB04040100), the National Basic Research Program of China (973
Program, Grants No. 2012CB922002 and No. 2011CBA00101), and the
Chinese Academy of Sciences.


\begin{references}
\bibitem{Hosono}
Y. Kamihara, T. Watanabe, M. Hirano, and H. Hosono, J. Am. Chem. Soc. \textbf{130}, 3296 (2008).

\bibitem{Chen}
X. H. Chen, T. Wu, G. Wu, R. H. Liu, H. Chen, and D. F. Fang, Nature
(London) \textbf{453}, 761 (2008).

\bibitem{BaK}
M. Rotter, M. Tegel, and D. Johrendt, Phys. Rev. Lett. \textbf{101},
107006 (2008).

\bibitem{Wang XF PRL}
X. F. Wang, T. Wu, G. Wu, H. Chen, Y. L. Xie, J. J. Ying, Y. J. Yan,
R. H. Liu, and X. H. Chen, Phys. Rev. Lett. \textbf{102}, 117005
(2009).

\bibitem{Hiroshi}
Kunihiro Kihou, Taku Saito, Shigeyuki Ishida, Masamichi Nakajima,
Yasuhide Tomioka, Hideto Fukazawa, Yoh Kohori, Toshimitsu Ito,
Shin-ichi Uchida, Akira Iyo, Chul-Ho Lee, Hiroshi Eisaki, J. Phys.
Soc. Jpn, \textbf{79}, 124713 (2010).

\bibitem{Wang XF}
J. S. Kim, E. G. Kim, G. R. Stewart, X. H. Chen and X. F. Wang,
Phys. Rev. \textbf{83}, 172502 (2011).

\bibitem{Ding H}
T. Sato, K. Nakayama, Y. Sekiba, P. Richard, Y. M. Xu, S. Souma, T.
Takahashi, G. F. Chen, J. L. Luo, N. L. Wang, and H. Ding, Phys.
Rev. Lett. \textbf{103}, 047002(2009).

\bibitem{Haas Alphen}
T. Terashima, M. Kimata, N. Kurita, H. Satsukawa, A. Harada, K.
Hazama, M. Imai, A. Sato, K. Kihou, C. H. Lee, H. Kito, H. Eisaki,
A. Iyo, T. Saito, H. Fukazawa, Y. Kohori, H. Harima, and S. Uji, J.
Phys. Soc. Jpn. \textbf{79}, 053702 (2010).

\bibitem{Chu CW}
Kalyan Sasmal, Bing Lv, Bernd Lorenz, Arnold M. Guloy, Feng Chen,
Yu-Yi Xue, and Ching-Wu Chu, Phys. Rev. Lett. \textbf{101}, 107007
(2008).

\bibitem{Li SY}
J. K. Dong, S. Y. Zhou, T. Y. Guan, H. Zhang, Y. F. Dai, X. Qiu, X.
F. Wang, Y. He, X. H. Chen, and S. Y. Li, Phys. Rev. Lett.
\textbf{104}, 087005 (2010).

\bibitem{Wang NL}
Taichi Terashima, Motoi Kimata, Hidetaka Satsukawa, Atsushi Harada,
Kaori Hazama, Shinya Uji, Hisatomo Harima, Gen-Fu Chen, Jian-Lin
Luo, and Nan-Lin Wang, J. Phys. Soc. Jpn. \textbf{78}, 063702
(2009).

\bibitem{penetration depth}
K. Hashimoto, A. Serafin, S. Tonegawa, R. Katsumata, R. Okazaki, T.
Saito, H. Fukazawa, Y. Kohori, K. Kihou, C. H. Lee, A. Iyo, H.
Eisaki, H. Ikeda, Y. Matsuda, A. Carrington, and T. Shibauchi, Phys.
Rev. B \textbf{82}, 014526 (2010).

\bibitem{NMR}
H. Fukazawa, Y. Yamada, K. Konda, T. Saito, Y. Kohori, K. Kuga, Y.
Matsumoto, S. Nakatsuji, H. Kito, P.M.Shirage, K.Kihou, N.
Takeshita, C. H. Lee, A. Iyo, and H. Eisaki, J. Phys. Soc. Jpn.
\textbf{79}, 083712 (2009).

\bibitem{Yu WQ}
S. W. Zhang, L. Ma, Y. D. Hou, J. Zhang, T. L. Xia, G. F. Chen, J.
P. Hu, G. M. Luke, and W. Yu, Phys. Rev. B \textbf{81}, 012503
(2010).

\bibitem{Louis Taillefer}
J.-Ph. Reid, M. A. Tanatar, A. Juneau-Fecteau, R. T. Gordon, S.
Ren\'{e} de Cotret, N. Doiron-Leyraud, T. Saito, H. Fukazawa, Y.
Kohori, K. Kihou, C. H. Lee, A. Iyo, H. Eisaki, R. Prozorov, and
Louis Taillefer, Phys. Rev. Lett. \textbf{109}, 087001 (2012).

\bibitem{Wang AF}
A. F. Wang, S. Y. Zhou, X. G. Luo, X. C. Hong, Y. J. Yan, J. J.
Ying, P. Cheng, G. J. Ye, Z. J. Xiang, S. Y. Li, and X. H. Chen,
arXiv: 1206.2030.

\bibitem{KNa}
M. Abdel-Hafiez, V. Grinenko, S. Aswartham, I. Morozov, M. Roslova,
O. Vakaliuk, S. L. Drechsler, S. Johnston, D. V. Efremov, J. Van den
Brink, H. Rosner, M. Kumar, C. Hess, S. Wurmehl, A.U.B. Wolter,
B.B\"{u}chner, E. L. Green, J. Wosnitza, P. Vogt, A. Reifenberger,
C. Enss, and R. Klingeler, arXiv: 1301.5257.

\bibitem{Science}
K. Okazaki, Y. Ota, Y. Kotani, W. Malaeb, Y. Ishida, T. Shimojima,
T. Kiss, S. Watanabe, C.-T. Chen, K. Kihou, C. H. Lee, A. Iyo, H.
Eisaki, T. Saito, H. Fukazawa, Y. Kohori, K. Hashimoto, T.
Shibauchi, Y. Matsuda, H. Ikeda, H. Miyahara, R. Arita, A. Chainani,
S. Shin, Science \textbf{337} 1314 (2012).

\bibitem{Ni N}
N. Ni, S. L. Budfko, A. Kreyssig, S. Nandi, G. E. Rustan, A. I.
Goldman, S. Gupta, J. D. Corbett, A. Kracher, and P. C. Canfield,
Phys. Rev. B \textbf{78}, 014507 (2008).

\bibitem{Canfield}
Sergey L. Bud'Ko, Yong Liu, Thomas A. Lograsso, and Paul C.
Canfield, Phys. Rev. B \textbf{86}, 224514 (2012).

\bibitem{BNC}
J. S. Kim, G. R. Stewart, S. Kasahara, T. Shibauchi, T. Terashima,
and Y. Matsuda, J. Phys. : Condens. Matter \textbf{23}, 222201
(2011).

\bibitem{magnetic susceptibility}
V. Grinenko, S. L. Drechsler, M. Abdel-Hafiez, S. Aswartham, A. U.
B. Wolter, S. Wurmehl, C. Hess, K. Nenkov, G. Fuchs, D. V. Efremov,
B. Holzapfel, J. van den Brink, and Buchner, Phys. Status Solidi B,
pssc. 201200805.R1 (2012).

\bibitem{d wave}
O. J. Taylor, A. Carrington, and J. A. Schlueter, Phys. Rev. Lett.
\textbf{99}, 057001 (2007).

\bibitem{a model}
H. Padamsee, J. Low Temp. Phys. \textbf{12}, 387 (1973).

\bibitem{Wang AF Na}
A. F. Wang, X. G. Luo, Y. J. Yan, J. J. Ying, Z. J. Xiang, G. J. Ye,
Z. Y. Li, W. J. Hu, and X. H. Chen, Phys. Rev. B \textbf{85}, 224521
(2012).

\bibitem{d wave2}
V. Z. Kresin and S. A. Wolf, Physica C \textbf{169}, 476 (1990).

\bibitem{LiSYC}
X. C. Hong, X. L. Li, B. Y. Pan, L. P. He, A. F. Wang, X. G. Luo, X.
H. Chen, and S. Y. Li, arXiv: 1302.2300.


\end{references}
\end{document}